\documentclass[%
reprint,
superscriptaddress,
 amsmath,amssymb,
 aps,
noeprints,
floats,
10pt]{revtex4-2}

\usepackage[utf8]{inputenc}
\usepackage{physics}
\usepackage{amsmath}
\usepackage{caption}
\usepackage{bm}
\usepackage{upgreek}
\usepackage{xr}
\usepackage[dvipsnames]{xcolor}
\usepackage{subcaption}
\usepackage{float}
\usepackage{tablefootnote}
\usepackage{footnote}
\usepackage{graphicx}
\usepackage{braket}
\usepackage[export]{adjustbox}
\usepackage{tabularx}
\usepackage{lipsum}  
\newcommand{\comment}[1]{}
\usepackage{multirow}
\usepackage{makecell}

\usepackage{calrsfs} 
\DeclareMathAlphabet{\pazocal}{OMS}{zplm}{m}{n}

\usepackage[normalem]{ulem}
\usepackage{color}
\usepackage{xcolor}

\definecolor{Mygrey}{gray}{0.80}
\definecolor{lteal}{rgb}{0.10,0.60,0.70}
\definecolor{dkred}{rgb}{0.60,0.10,0.00}
\definecolor{Navy}{rgb}{0.00,0.00,0.60}
\definecolor{Magenta}{rgb}{0.94,0.20,0.90}
\definecolor{Green}{rgb}{0.24,0.71,0.29}
\definecolor{Teal}{rgb}{0.00,0.50,0.50}

\begin{document}
\pagestyle{plain}

\title{Foundations of the ionization potential condition for localized electron removal in density functional theory}

\author{Guy Ohad}
    \affiliation{Department of Molecular Chemistry and Materials Science, Weizmann Institute of Science, Rehovoth 76100001, Israel}
\author{Mar\'ia Camarasa-G\'omez}
    \affiliation{Department of Molecular Chemistry and Materials Science, Weizmann Institute of Science, Rehovoth 76100001, Israel}
    \affiliation{Centro de F\'isica de Materiales (CFM-MPC), CSIC-UPV/EHU, Paseo Manuel de Lardizabal 5, 20018 Donostia-San Sebastián, Spain}
\author{Jeffrey B. Neaton}
    \affiliation{Department of Physics, University of California, Berkeley, Berkeley, California 94720, USA}
    \affiliation{Materials Sciences Division, Lawrence Berkeley National Laboratory, Berkeley, California 94720, USA}
    \affiliation{Kavli Energy NanoSciences Institute at Berkeley, University of California, Berkeley, Berkeley, California 94720, USA}
\author{Ashwin Ramasubramaniam}
    \affiliation{Department of Mechanical and Industrial Engineering, University of Massachusetts Amherst, Amherst, MA 01003, USA}
    \affiliation{Materials Science and Engineering Graduate Program, University of Massachusetts Amherst, Amherst, MA 01003, USA}
\author{Tim Gould}
    \affiliation{Queensland Micro- and Nanotechnology Centre, Griffith University, Nathan, QLD 4111, Australia}
\author{Leeor Kronik}
   \affiliation{Department of Molecular Chemistry and Materials Science, Weizmann Institute of Science, Rehovoth 76100001, Israel}

\begin{abstract}
Optimal tuning of functional parameters in density functional theory approximations, based on enforcing the ionization potential theorem, has emerged as the method of choice for the non-empirical prediction of the electronic structure of finite systems. This method has recently been extended to the bulk limit, based on an \textit{ansatz} that generalizes the ionization potential theorem to the removal of an electron from a localized Wannier orbital. This Wannier-localization based optimal tuning method has been shown to be highly successful for a wide range of periodic systems, accurately predicting electronic and optical properties. However, a rigorous theoretical justification for its foundational \emph{ansatz} has been lacking. Here, we establish an ionization potential condition for the removal of an electron from a Wannier-localized orbital, by extending the piecewise linearity and Janak's theorems in density functional theory. We also provide numerical evidence supporting our theory.    
\end{abstract}

\maketitle
\thispagestyle{plain}

Density functional theory (DFT) has become the workhorse of first principles calculations across an unusually wide array of disciplines, as it offers predictive insights into many properties of molecules and solids, while offering an excellent balance between computational cost and accuracy \cite{teale_2022_dft_perspective,Sholl,cohen_louie_condensed_matter_book, martinElectronicStructureBasic2020}. However, DFT has historically struggled, often even qualitatively, in the description of electron and optical spectroscopy. Specifically, the band gap of semiconductors and insulators has been thought to be outside the reach of DFT even in principle, and the associated optical absorption spectrum outside the reach of time-dependent DFT with standard density functional approximations (DFAs) \cite{kuemmel_kronik_2008, onida2002electronic, bryenton2023delocalization,scuseria2021advancing}.

In molecules, predictions of electronic and optical excitations have been found to be significantly improved \cite{bryenton2023delocalization,stein_kronik_baer_2009, dabo2010koopmans, lany2009polaronic,miceli_pasquarello_2018,linscott2023koopmans, trushin2025improving}
by forcing DFAs to obey an exact condition -- the ionization potential (IP) theorem \cite{perdew_balduz_1982, Almbladh_von_Barth_1985, yang2000degenerate, cohenFractionalChargePerspective2008}. It states that for the exact functional, the IP from total energies difference is equal and opposite to the eigenvalue associated with the highest-occupied orbital of the (generalized) Kohn-Sham [(G)KS] \cite{kohn_sham_1965, seidl_levy_1996} system. This relation is expressed as
\begin{equation}
    \label{eq:IPT}
    E^{N-1}_\text{g.s.} - E^N_\text{g.s.} = -\epsilon_H,
\end{equation}
where $E^N_\text{g.s.}$ and $E^{N-1}_\text{g.s.}$ are the ground state (g.s.) energies of the $N$ and $N-1$ electron systems, respectively, and $\epsilon_H$ is the highest-occupied eigenvalue of the (G)KS effective Hamiltonian. The IP theorem can be proven by combining the piecewise linearity (PWL) of the total energy in the fractional electron number between integer points \cite{perdew_balduz_1982,yang2000degenerate, kronik_kummel_2020} and Janak's theorem \cite{janak_1978}. Importantly, Eq.~\eqref{eq:IPT} must be obeyed for exact (G)KS DFT, but is typically violated by DFAs \cite{mori-sanchez_yang_2008, kronik_kummel_2020}.

Over the last two decades, multiple approaches have been developed to enforce the IP theorem (or a generalization thereof) in DFAs, in order to improve electronic structure predictions \cite{stein_kronik_baer_2009, dabo2010koopmans, lany2009polaronic,miceli_pasquarello_2018,linscott2023koopmans, trushin2025improving}. Here, as an important example we focus primarily on optimal tuning (OT) \cite{stein_kronik_baer_2009} of (screened) range-separated hybrid [(S)RSH] functionals \cite{savin1995density, yanaiNewHybridExchange2004, kronik_stein_refaely-abramson_baer_2012, refaely-abramson_kronik_2013}, where functional parameters are selected such that Eq.~\eqref{eq:IPT} is satisfied in the functional. This is because the exceptional accuracy of OT-RSH in predicting electronic and optical excitations in molecules has been repeatedly demonstrated (see, e.g., Refs.~\cite{salzner2009koopmans,stein_baer_2010, refaely_kronik_2011, kronik_stein_refaely-abramson_baer_2012, autschbach_srebro_2014, phillips_dunietz_2014, foster_allendorf_2014, korzdorfer_bredas_2014,tamblyn2014simultaneous, faber2014excited, alipour2017shedding}) and theoretically justified \cite{stein_baer_2012}. However, the argumentation is broadly applicable. 

Unfortunately, it was recognized early on that in the bulk limit, \emph{any} DFA exhibits PWL and thus trivially satisfies the IP theorem, regardless of accuracy (or lack thereof) in the resulting electronic structure \cite{mori-sanchez_yang_2008, kraisler_kronik_2014, vlcek_eisenberg_steinle-neumann_baer_2015, gorling_2015}. As a result, the predictive power of IP-theorem-based approaches, including OT, is lost for periodic systems. This deficiency is a result of the delocalized nature of the (G)KS orbitals. A popular heuristic solution has been to instead work with localized-orbital based approaches, giving rise to a wide range of broadly applicable methods to improve electronic structure predictions in the bulk limit within DFT (see, e.g., Refs.~\cite{wing_2021,chan_ceder_2010,li2015local,nguyen_marzari_2018,degennaro2022bloch,linscott2023koopmans, ma_wang_2016,liLocalizedOrbitalScaling2018, mahler2022localized, miceli_pasquarello_2018, chen_pasquarello_2018, yang2022one, BLOR2023, BLOR2024}).
However, a rigorous theoretical justification of these approaches has been lacking -- a critical gap that we remedy in this Letter by generalizing the IP theorem to address localized removal of an electron. 


One particular approach that uses localized orbitals is the Wannier-localization based optimal tuning (WOT) method \cite{wing_2021}, where the parameters of SRSH are selected based on enforcing an \textit{ansatz} generalization of the IP theorem, inspired by the work of Ma and Wang \cite{ma_wang_2016}. This condition is expressed as
\begin{equation}
    \label{eq:wot_ansatz}
    \tilde{E}^{N-1} - E^N_\text{g.s.} = -\tilde{\epsilon}_H^N,
\end{equation}
where $\tilde{E}^{N-1}$ is the total energy of an $(N-1)$-electron system under the constraint that the electron is removed from the Wannier orbital associated with the occupied manifold having the maximal expectation energy, $\tilde{\epsilon}_H$, with respect to the GKS Hamiltonian. WOT-SRSH has proven to be highly accurate in predicting electronic and optical excitations for a wide range of periodic systems, both in itself and as a starting point for many-body perturbation theory calculations \cite{wing_2021,gant2022optimally,ohad_2022_wotsrsh_haps,ohad_gant_2023_metal_oxides,ke_2024_first,ohad_2024_chains,sagredo2024electronic,camarasa_2024_vdw_wot, florio_2025_pbse2}. In principle, once the WOT-SRSH parameters are determined, the functional can be used to calculate any property of interest. Previous work has mainly focused on fundamental band gaps -- calculated from the difference between the eigenvalues that correspond to the valence band maximum and conduction band minimum, as well as optical absorption spectra, obtained from linear-response time-dependent DFT. However, applicability extends beyond these properties. For example, it has been shown that the optimal parameters determined from the bulk are also useful quantitatively for studies of surfaces \cite{sagredo_2025_surfaces} and defects \cite{ke_2025_defects}. We surmise that other properties - from dielectric response to magnetic moments - can also be addressed.

Given the success of WOT-SRSH, an obvious question is \textit{why} this approach yields such accurate results. However, to date, the WOT \textit{ansatz} of Eq.~\eqref{eq:wot_ansatz} (or an approximate form of it) has not been derived from fundamental principles. In this Letter, we remedy this deficiency by validating the WOT \textit{ansatz} from a theoretical standpoint, based on suitable extensions of the proof of the original IP theorem. Specifically, we provide a localized electron removal PWL condition and derive an exact generalized Janak's theorem, based on considerations involving unitarily transformed orbitals. The combination of these extensions gives rise to a generalized IP theorem for a removal of a localized electron, that explains the success of existing heuristic approaches and points the way to further developments. We further provide numerical evidence in support of the generalized PWL condition and its relation to the IP \textit{ansatz}.

Our first step is to establish a generalized PWL condition for localized electron removal, noting that the density resulting from a local removal of an electron in an extended system does not correspond to the ground-state density of the $(N-1)$-electron system. To illustrate this important point, consider first the hole associated with the cation of a finite chain of $M$ repeated units (e.g., as in an oligomer). In general, the hole is spread out over all $M$ units and may be thought of as resulting from the removal of $1/M$ of an electron from each of $M$ localized orbitals centered on each unit. If we assume that the localized frontier orbitals are energetically well-separated from other orbitals, it follows that the lowest $M-1$ excited states of the chain cation also involve removal of $1/M$ electrons from the localized frontier orbitals, but with different phases on each unit. That is, the $M$ lowest lying eigenstates of the cation may be obtained from superpositions of localized states. The same analysis may, of course, be applied to any periodic system, where $M \rightarrow \infty$ in principle, but can be finite in practice if the localization is performed in a supercell.

Crucially, it follows from basic linear algebra that the localized states may equivalently be obtained by superpositions of the $M$ eigenstates. Thus, rather than restricting to the usual ground state analysis, we start our analysis by considering an \textit{ensemble} \cite{GOK1, GOK2, GOK3} that mixes $M$ ground- and excited states -- defined via an equally weighted ensemble density matrix (EDM) operator,
\begin{equation}
    \label{edm_N-1}
    \hat{\Gamma}^{N-1}_e \equiv \frac{1}{M} \sum_{i=1}^M \ket{\Psi_i^{N-1}} \bra{\Psi_i^{N-1}}\;.
\end{equation}
Here, $\qty{\ket{\Psi_i^{N-1}}}$ are eigenstates of the many-body Hamiltonian, $\hat{H}$, the sum is over the subspace of the $M$ lowest-energy ones, and spin indices are suppressed throughout for simplicity.
The ensemble energy is then given by
\begin{equation}
    \label{eq:ensemble_energy_def}
    \pazocal{E}^{N-1}_e = \textrm{Tr} \left[ \hat{\Gamma}^{N-1}_e \hat{H} \right]=\frac{1}{M}\sum_{i=1}^M E^{N-1}_i,
\end{equation}
and the ensemble density by
\begin{equation}
    \label{eq:ensemble_density_def}
    n^{N-1}_e = \textrm{Tr} \left[ \hat{\Gamma}^{N-1}_e \hat{n} \right]=\frac{1}{M}\sum_{i=1}^M n^{N-1}_i.
\end{equation}
Here, $E^{N-1}_i \equiv \bra{\Psi_i^{N-1}}\hat{H}\ket{\Psi_i^{N-1}}$ and $n^{N-1}_i \equiv \bra{\Psi_i^{N-1}}\hat{n}\ket{\Psi_i^{N-1}}$, where $\hat{n}$ is the density operator, and $\textrm{Tr}$ stands for trace. Because it represents an equally weighted ensemble, the EDM in Eq.~\eqref{edm_N-1} remains invariant under a unitary transform of $\qty{\ket{\Psi_i^{N-1}}}$. Thus, the set of wavefunctions
\begin{equation}
    \ket{\tilde{\Psi}_p^{N-1}} \equiv \sum_{i=1}^M \pazocal{U}_{ip} \ket{\Psi_i^{N-1}},
    \label{eq:Unitary}
\end{equation}
for an arbitrary $M \cross M$ unitary matrix $\pazocal{U}$, obeys
\begin{equation}
    \label{eq:edm_invariance_to_u}
    \hat{\Gamma}^{N-1}_e = \frac{1}{M} \sum_{p=1}^M \ket{\tilde{\Psi}_p^{N-1}} \bra{\tilde{\Psi}_p^{N-1}}.
\end{equation}
That is, the ensemble can be described as a weighted sum of true eigenstates, or equivalently as a weighted sum of appropriate superpositions of eigenstates.

Let us now choose $\pazocal{U}$ such that: (a) The density of the ``hole'' associated with each state $\ket{\tilde{\Psi}_p^{N-1}}$, i.e., $n^N - \tilde{n}_p^{N-1}$, where $\tilde{n}_p^{N-1} \equiv \bra{\tilde{\Psi}_p^{N-1}}\hat{n}\ket{\tilde{\Psi}_p^{N-1}}$, is well-localized in the vicinity of some specific unit $p$, and is similar, up to translation, to the hole density of any other unit $p'$; and (b) the energies $\tilde{E}^{N-1}_p \equiv \bra{\tilde{\Psi}_p^{N-1}}\hat{H}\ket{\tilde{\Psi}_p^{N-1}}$ are all degenerate and thus equal to the average ensemble energy, $\tilde{E}^{N-1}_p = \tilde{E}^{N-1}_{p'} = ... = \pazocal{E}^{N-1}_e$. The modest assumption behind satisfying both (a) and (b) follows from physical intuition based on nearsightedness~\cite{Kohn1996-NS} and practical experience in systems of repeating units -- see Appendix A for details. Assumption (b), in fact, can always be satisfied -- see Appendix B for a proof. The numerical results presented below further validate the assumptions.

The localized removal states $\{\ket{\tilde{\Psi}_p^{N-1}}\}$ then describe a system where a localized electron removal from any of the $M$ units is equivalent to removal from any of the other units (up to edge effects which we assume to be negligible, and do not occur in fully periodic systems).
This construction can be thought of as the many-body analogy of removal of a maximally localized Wannier orbital \cite{marzari_vanderbilt_2012}, with the localized object being the many-body hole. In the case of the chain scenario discussed above, this is akin to applying the orbital-based argument to the $M$ lowest-lying many-body wavefunctions. In larger or bulk systems it can be applied to a supercell of $M$ units that is sufficiently large to represent the bulk of the $(N-1)$-electron system, and to allow for a state to be localized within it \footnote{Electron removal from a supercell with $M$ units is equivalent to removal using $M$ reciprocal space $\bm{k}$-points with a unit cell - see, e.g., Ref.\ \cite{vlcek_eisenberg_steinle-neumann_baer_2015} for a detailed discussion.}.
We note that in the above we have, for simplicity, treated the number of repeating cells used and number of excited states as one and the same. In general, the latter is expected to be an integer multiple of the former, e.g., to account for all pertinent states in the valence band. In practice, the number of states $M$ should be sufficiently large to localize the many-body holes. We also emphasize that a repeating unit in our context is not necessarily confined to the smallest possible unit (namely, a primitive unit cell), but can be in itself composed of several primitive units.

Having introduced ensemble considerations above, we now extend our analysis to the case of fractional number of electrons. We \emph{define} the relevant $(N-q)$-electron system ($0 < q < 1$) to be a fractional-charge ensemble \cite{perdew_balduz_1982}, $\hat{\gamma}_f^{N-q}=(1-q)\ket{\varphi^N}\bra{\varphi^N} + q\hat{\gamma}^{N-1}_e$, composed of an arbitrary $N$-electron state and an arbitrary $(N-1)$-electron excited state ensemble, $\hat{\gamma}^{N-1}_e$, formed on $M$ arbitrary orthonormal wavefunctions $\{\varphi^{N-1}\}$.
The ground-state EDM is found by solving $\min_{\varphi^N,\{\varphi^{N-1}\}}\text{Tr}[\hat{\gamma}_f^{N-q}\hat{H}]\equiv \text{Tr}[\hat{\Gamma}^{N-q}_f\hat{H}]$, which, using standard ensemble argumentation \footnote{To show this result, first recognize that the $N$- and $(N-1)$-electron states may be minimized separately due to differing electron numbers. Then, the Hohenberg-Kohn principle yields $\varphi^N\to \Psi_{\text{g.s.}}^N$ and the Gross-Olivera-Kohn principle yields {Eq.~\eqref{edm_N-1}}.}, leads to
\begin{equation}
    \label{edm_N-q}
    \hat{\Gamma}^{N-q}_f \equiv (1-q) \ket{\Psi^N_\textrm{g.s.}} \bra{\Psi^N_\textrm{g.s.}} + q \hat{\Gamma}^{N-1}_e.
\end{equation}
This fractional electron EDM is composed of the unique $N$-electron ground state and the lowest $M$ eigenstates of the $(N-1)$-electron Hamiltonian.
It follows, using Eqs.~\eqref{eq:ensemble_energy_def} and \eqref{eq:ensemble_density_def}, that
\begin{equation}
    \begin{aligned}
    n^{N-q}_f & = \textrm{Tr} \left[ \hat{\Gamma}^{N-q}_f \hat{n} \right] 
    = (1-q)n^N_\textrm{g.s.} + q n^{N-1}_e,
    \label{eqn:nNq}
    \end{aligned}
\end{equation}
and
\begin{equation}
    \begin{aligned}
    \pazocal{E}^{N-q}_f & = \textrm{Tr} \left[ \hat{\Gamma}^{N-q}_f \hat{H} \right] 
    = (1-q)E^N_\text{g.s.} + q \pazocal{E}^{N-1}_e,
    \label{eqn:ENq}
    \end{aligned}
\end{equation}
are the density and energy, respectively, of the $(N-q)$-electron system. In light of the invariance of the excited-state EDM to a unitary transformation [Eq.~\eqref{eq:edm_invariance_to_u}], Eqs.~\eqref{eqn:nNq} and \eqref{eqn:ENq} are also left unchanged under it.
However, by exploiting (a) we obtain
\begin{equation}
    \label{eq:pwl_local_density}
    \tilde{n}_p^{N-q} = (1-q)n^N_\textrm{g.s.} +q\tilde{n}_p^{N-1},
\end{equation}
for any $1 \leq p \leq M$, as a localized condition that is equivalent to Eq.~\eqref{eqn:nNq}. Likewise, Assumption (b) yields
\begin{equation}
    \label{eq:pwl_local_energy}
    \pazocal{E}^{N-q}_f = (1-q) E^N_\text{g.s.} + q \tilde{E}^{N-1}_p.
\end{equation}
Eqs.~\eqref{eq:pwl_local_density} and \eqref{eq:pwl_local_energy} thus constitute PWL conditions for the density and energy associated with \emph{localized} electron removal from any of the units, and are the first major result of this Letter.

We now proceed to derive a generalized Janak's theorem. We start from the pure-state total energy functional, which in KS theory is usually written in the form \cite{kohn_sham_1965}
\begin{equation}
    \label{eq:toten}
    E[n] = T_s[n] + E_\textrm{ext}[n] + E_{H}[n] + E_\textrm{xc}[n],
\end{equation}
where $T_s[n]$ is the kinetic energy of the non-interacting KS system, $E_\textrm{ext}[n]$ is the energy due to the external potential, $E_{H}[n]$ is the Hartree energy, and $E_\textrm{xc}[n]$ is the exchange-correlation energy. The KS equation is
\begin{equation}
    H_\textrm{KS} \psi_i(\bm{r}) = \epsilon_i \psi_i(\bm{r}),
\end{equation}
where $\psi_i$ and $\epsilon_i$ are the KS orbitals and eigenvalues, respectively, and
\begin{equation}
    H_\textrm{KS} \equiv -\frac{1}{2}\nabla^2 + v_\textrm{ext}(\bm{r})+v_H(\bm{r})+v_\textrm{xc}(\bm{r}),
\end{equation}
with the subscripts of the potential terms corresponding to the subscripts of the energy terms.


The density and kinetic energy of a fractionally occupied KS system, as defined by Janak \cite{janak_1978}, are
$n (\bm{r}) = \sum_i f_i \abs{\psi_i(\bm{r})}^2$ and
$T_s  = \sum_i f_i \bra{\psi_i} -\tfrac{1}{2}\nabla^2 \ket{\psi_i}$,
where $\qty{f_i}$ are occupation numbers between 0 and 1. Within these definitions, Janak's theorem states that
\begin{equation}
    \label{eq:janaks_theorem}
    \frac{\partial E}{\partial f_{i}} = \epsilon_{i}.
\end{equation}
Now consider a unitary transform of the KS orbitals
\begin{equation}
    \label{eq:psi_tilde}
    \ket{\tilde{\psi}_i} = \sum_j U_{ji} \ket{\psi_j},
\end{equation}
where $U_{ji}$ are the matrix elements of an arbitrary unitary matrix. The density and kinetic energy can then be expressed in terms of the unitarily transformed orbitals as 
\begin{equation}
    \begin{aligned}
        n (\bm{r}) & = \sum_j \sum_{j'} \tilde{f}_{jj'} \tilde{\psi}_{j'}^*(\bm{r}) \tilde{\psi}_{j}(\bm{r})\\
        T_s & = \sum_j \sum_{j'} \tilde{f}_{jj'} \bra{\tilde{\psi}_{j'}} -\tfrac{1}{2}\nabla^2\ket{\tilde{\psi}_{j}},
    \end{aligned}
\end{equation}
where the occupancy matrix elements $\tilde{f}_{jj'}$ are defined as
\begin{equation}
    \tilde{f}_{jj'} \equiv \sum_i f_i U_{ij'} U_{ij}^*.
\end{equation}
The trace of the occupancy matrix is invariant to the unitary transform, allowing for a (possibly fractional) number of electron counted through summation over diagonal elements only, namely $\sum_i f_i=\sum_j \tilde{f}_{jj}$. Note that the unitary transform guarantees that $0 \leq \tilde{f}_{jj} \leq 1$.

Next, consider the variation of the total energy with respect to one of the $\tilde{f}_{ii'}$ (and independent of the other occupancies), allowing for orbital relaxation. We obtain
\begin{equation}
\label{eq:chain_rule_dE_df}
\begin{split}
\begin{aligned}
    \frac{\partial E}{\partial\tilde{f}_{ii'}}  = & \sum_{j}\frac{\partial E}{\partial f_{j}}\frac{\partial f_{j}}{\partial\tilde{f}_{ii'}}+\sum_{j}\int\frac{\delta E}{\delta\psi_{j}^{*}\left(\boldsymbol{r}\right)}\frac{\partial\psi_{j}^{*}\left(\boldsymbol{r}\right)}{\partial\tilde{f}_{ii'}}d^{3}r \\
    & + \sum_{j}\int\frac{\delta E}{\delta\psi_{j}\left(\boldsymbol{r}\right)}\frac{\partial\psi_{j}\left(\boldsymbol{r}\right)}{\partial\tilde{f}_{ii'}}d^{3}r \\
     = & \tilde{\epsilon}_{ii'}+\sum_{j}\epsilon_{j}\frac{\partial}{\partial\tilde{f}_{ii'}}\left[\int\psi_{j}\left(\boldsymbol{r}\right)\psi_{j}^{*}\left(\boldsymbol{r}\right)d^{3}r\right],
\end{aligned}
\end{split}
\end{equation}
where $\tilde{\epsilon}_{ii'} \equiv \bra{\tilde{\psi}_{i'}}\hat{H}_\textrm{KS}\ket{\tilde{\psi}_{i}}$. Because $\qty{\psi_j}$ are normalized, the second term vanishes and we get
\begin{equation}
    \label{eq:generalized_janak}
    \frac{\partial E}{\partial\tilde{f}_{ii'}} = \tilde{\epsilon}_{ii'}.
\end{equation}
Eq.~\eqref{eq:generalized_janak} can be thought of as a generalized Janak's theorem, with Janak's original theorem being a special case thereof, where the unitary matrix is the identity matrix. Eq.~\eqref{eq:generalized_janak} is the second major result of this Letter. Similar to Janak's theorem, no assumptions have been made about the exchange-correlation energy, implying that the result is applicable to both the exact functional and approximate ones. Additionally, we point out that an extension of the steps above to a global hybrid functional GKS system is simple and only requires accounting for the fractional occupancies in the exact exchange energy via \footnote{For pedagogical reasons, the KS case and the hybrid GKS case are discussed here separately. The two cases, however, may be addressed simultaneously by defining a one-body reduced density matrix $\rho (\bm{r},\bm{r}') \equiv \sum_i f_i \psi_i^*(\bm{r}')\psi_i(\bm{r})=\sum_j\sum_{j'} \tilde{f}_{jj'} \tilde{\psi}_{j'}^*(\bm{r}')\tilde{\psi}_j(\bm{r})$, in which case the total energy of Eq.~\eqref{eq:toten} can be written as a unique functional of $\rho$. Then, $E_\textrm{xc}[\rho]$ captures the usual mixture of exact and approximate exchange and correlation of a hybrid functional.}
\begin{equation}
\label{eq:exactexchange}
    \begin{split}
    \begin{aligned}
    E_{\text{x}} = -& \frac{1}{2} \sum_{i} \sum_{j} f_i f_j \int \int \psi_i^*(\bm{r}) \psi_j(\bm{r}) \\ & \cross \frac{1}{\abs{\bm{r} - \bm{r}'}} \psi_i(\bm{r}') \psi_j^*(\bm{r}') d^3r d^3r'\\
    = -& \frac{1}{2} \sum_{i} \sum_{i'} \sum_{j} \sum_{j'} \tilde{f}_{ii'} \tilde{f}_{jj'} \int \int \tilde{\psi}_{i'}^*(\bm{r}) \tilde{\psi}_j(\bm{r}) \\ & \cross \frac{1}{\abs{\bm{r} - \bm{r}'}}  \tilde{\psi}_i(\bm{r}') \tilde{\psi}_{j'}^*(\bm{r}') d^3r d^3r',
    \end{aligned}
    \end{split}
\end{equation}
and similarly for RSH functionals by including, e.g., an error function within the integrals \cite{kronik_stein_refaely-abramson_baer_2012}. 

We are now ready to use the above results to establish a generalized IP condition for a local removal. As above, we make some physical assumptions about locality but otherwise use exact results.
Consider a unitary matrix $U$ for which there exists a subset $\qty{\tilde{\psi}_p}$ in the $N$-electron system that is well-localized on specific units such that the densities $\abs{\tilde{\psi}_p (\bm{r})}^2$ are all similar, up to translation, and are a good approximation to the hole densities obtained from $\pazocal{U}$ of the many-body system.  In other words, we assume that $\pazocal{U}$ is the many-body equivalent of $U$, only that for the many-body case it produces localized holes and not orbitals. We can then assume that the density of a system where one of the $\qty{\tilde{\psi}_p}$ is fractionally occupied is a good approximation for the density of the system described by Eqs.~\eqref{eq:pwl_local_density} and \eqref{eq:pwl_local_energy}. We now consider a variation in the total number of electrons by allowing a change only in the occupancy associated with the maximally occupied $\tilde{\epsilon}_{ii}$, namely $\tilde{f}_{H} \equiv 1-q$ (recall that electrons are counted only through diagonal elements, $i=i'=H$). We can then combine result \eqref{eq:pwl_local_energy} with result \eqref{eq:generalized_janak} to obtain
\begin{equation}
    \label{eq:generalized_IPT}
    \tilde{E}^{N-1} - E^N_\text{g.s.} = -\tilde{\epsilon}_{H}^N.
\end{equation}
Eq.~\eqref{eq:generalized_IPT} is a generalized localized-orbital-occupation based IP `theorem', which provides a rigorous justification of the IP \textit{ansatz} of Eq.~\eqref{eq:wot_ansatz}. It is the third major, and key, result of this study. Importantly, it is a significant generalization of the original IP theorem to the removal of an electron from a \emph{Wannier-localized} orbital, thus providing a rigorous theoretical framework for optimal tuning in the thermodynamic limit, in which conventional optimal tuning is unhelpful \cite{mori-sanchez_yang_2008}. We reiterate that the only formal assumptions made are: 1) that, for the $(N-1$)-electron system, we can obtain $M$ localized and nearly-degenerate states from a unitary transform of the lowest $M$ excited states; 2) that there is a Wannier-like transformation of \emph{non-interacting} (G)KS orbitals that can yield the localized hole densities of the \emph{interacting} system. Within these modest assumptions, Eq.~\eqref{eq:generalized_IPT} is an exact result.

We now wish to further discuss the link between the results presented above and the WOT strategy. Notably, the steps above apply to any choice of $U$ and the Wannier transform is included within the set of possible unitary transforms described by Eq.~\eqref{eq:psi_tilde}. If the unitary matrix, $U$, is the one associated with maximally localized Wannier orbitals, Eq.~\eqref{eq:generalized_IPT} reduces to the WOT \textit{ansatz}. An intuitive, though informal, claim can be proposed as follows: The closer $U$ is to the maximally localized unitary matrix, the greater the uniqueness and sensitivity achieved in bulk tuning. Conversely, the closer $U$ is to the identity matrix, the greater the nonuniqueness and insensitivity observed in bulk tuning. Furthermore, in a system of repeating units, maximally localized Wannier orbitals can be typically constructed such that there exists a subset that is localized on several of the different units with the same orbital nature and same energy. Removal of an electron from each of these Wannier orbitals is then consistent with the theory. 

\begin{figure}[htbp!]
\begin{centering}
\includegraphics[width=0.9\linewidth]{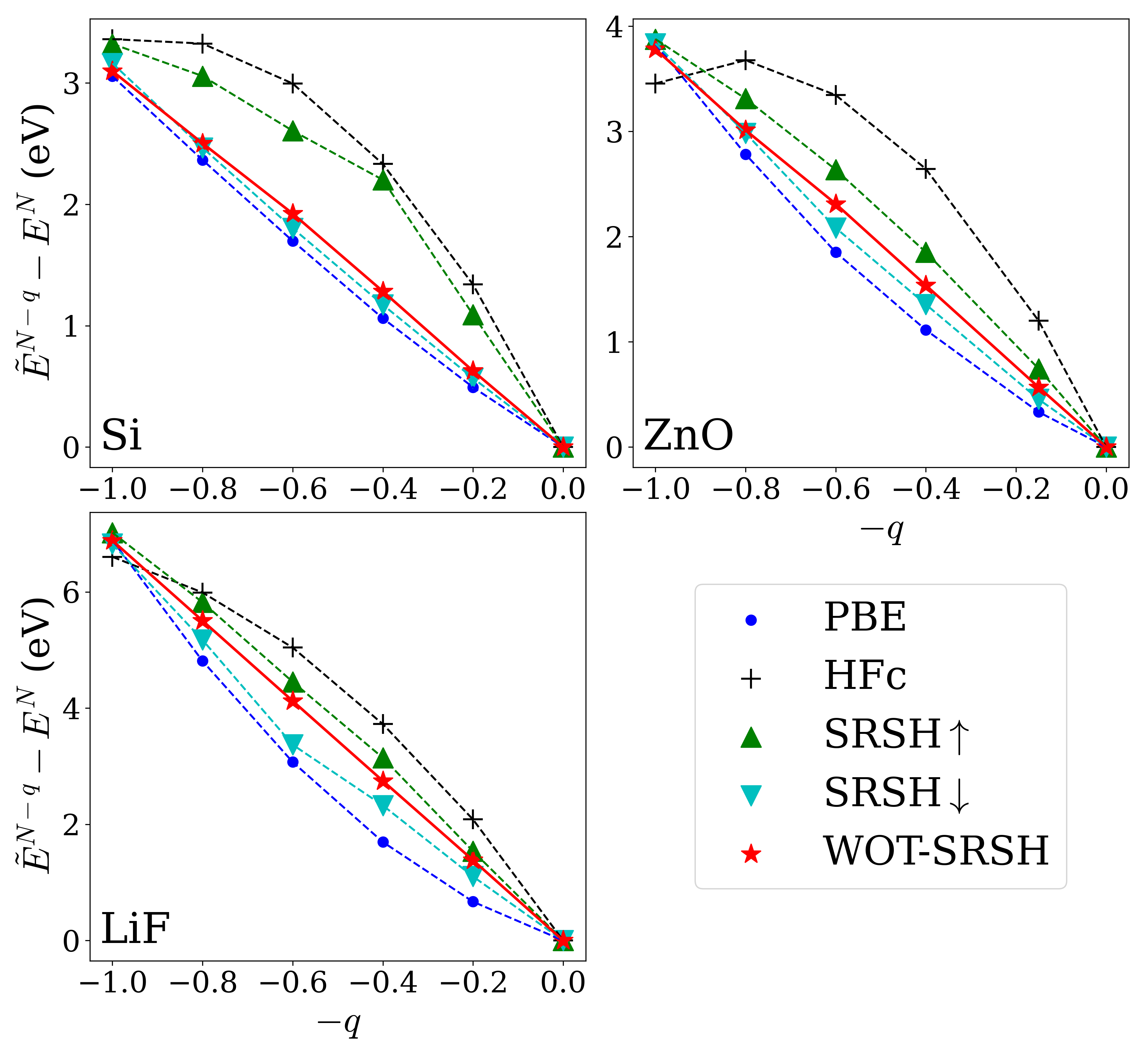}
    \caption{Total energy as a function of the fraction of Wannier orbital removed, for Si, ZnO, and LiF. The DFAs considered are PBE, Hartree-Fock plus semilocal correlation (HFc), SRSH that overestimates the band gap (SRSH$\uparrow$), SRSH that underestimates the band gap (SRSH$\downarrow$), and WOT-SRSH.}
    \label{fig:toten_vs_wann_occ}
\end{centering}
\end{figure}

To assess the reliability of our theory and justify its underlying assumptions, we turn to numerical results that demonstrate the PWL of WOT and its connection to the IP \textit{ansatz} in practical calculations. For that purpose, we define a procedure for removing a fraction of an electron from the highest-energy Wannier orbital, based on constraining its occupancy and enforcing its orthogonality to the other orbitals self-consistently \footnote{In these calculations we use the exact exchange expression of Eq.\ \eqref{eq:exactexchange} as is, i.e., without ensemble corrections. This is because the derivative discontinuity that is missing owing to the lack of ensemble corrections is what causes the curvature. The optimal parameter is obtained when there is no curvature, i.e., there is no derivative discontinuity and there is no need for ensemble corrections. See Ref.\ \cite{stein_baer_2012} for a detailed discussion.  We also note that the calculations are carried out in a supercell to minimize periodic image-charge interactions. See the SM for further details \cite{SM}.}. Further details on this numerical procedure and other computational parameters are given in the Supplemental Material (SM) \cite{SM}. Using this procedure, we exploit fractional Wannier calculations, as shown in Fig.~\ref{fig:toten_vs_wann_occ} and Table \ref{tab:wannier_curvature}, for Si, ZnO, and LiF. These materials were chosen to span a range of band gap values, from relatively narrow to very wide. The WOT-SRSH parameters for these materials have been determined in prior work \cite{wing_2021},\cite{ohad_gant_2023_metal_oxides},\footnote{The small difference between the band gap values reported here and the ones reported in Ref.~\cite{wing_2021} stems from the use of a different code, specifically from the difference between using norm-conserving pseudopotentials (Ref. ~\cite{wing_2021}) and the projector augmented wave method (this work).}. 

The PWL of $\tilde{E}$ with respect to $q$ within WOT-SRSH is clearly observed, as indicated by the small curvature, compared to the other DFAs which exhibit deviation from PWL. The very small remaining curvature is likely because the IP theorem and PWL are the same only to second order in the fractional charge within optimally-tuned approximations
– an issue previously observed within the context of the original IP theorem \cite{stein_baer_2012}.

We now focus on four other DFAs. First, the two limiting cases of the semilocal Perdew-Burke-Ernzerhof (PBE) functional \cite{perdew_burke_ernzerhof_1996} and Hartree-Fock plus PBE correlation (HFc) are considered. Second, we select non-optimally-tuned SRSH parameters such that the functional overestimates (SRSH$\uparrow$) and underestimates (SRSH$\downarrow$) the band gap (the parameters used are given in the SM \cite{SM}). These results demonstrate trends similar to those in unconstrained OT for small molecules \cite{stein_baer_2012, srebro_autschbach_2012_tuned}: The correct gaps are associated with a PWL curve, underestimated gaps are associated with convex curves, and overestimated gaps are associated with concave curves. This analysis confirms that WOT-SRSH successfully enforces PWL for localized electron removal. Furthermore, this establishes a clear connection between the functional’s band gap prediction and its deviation from PWL, via the generalized Janak theorem. The observed trends across different DFAs further support the validity of using PWL as a diagnostic for functional accuracy.

\begin{table}[htbp!]
  \centering
  \caption{Fundamental band gap, $E_g$, and curvature, $C$, determined from a second-order polynomial fit of the curves in Fig.~\ref{fig:toten_vs_wann_occ}, in eV.
  }
  \label{tab:wannier_curvature}
  \begin{ruledtabular}\begin{tabular}{|c|cc|cc|cc|}
    & \multicolumn{2}{c|}{Si} & \multicolumn{2}{c|}{ZnO} & \multicolumn{2}{c|}{LiF} \\
    \hline
        & $E_g$ & $C$ & $E_g$ & $C$ & $E_g$ & $C$ \\
    \hline
    PBE & 0.6 & 0.6 & 0.8 & 1.8 & 9.1 & 4.4 \\
    \hline
    HFc & 6.0 & -4.1 & 11.1 & -5.3 & 21.6 & -4.5 \\
    \hline
    SRSH$\uparrow$ & 3.6 & -3.0 & 5.6 & -1.3 & 17.9 & -1.2 \\
    \hline
    SRSH$\downarrow$ & 1.0 & 0.4 & 2.2 & 0.8 & 12.9 & 2.3 \\
    \hline
    WOT-SRSH & 1.2 & -0.2 & 3.5 & -0.1 & 15.3 & 0.0 \\
    \hline
    Experiment & 1.2 \footnote{Sum of the experimental room temperature fundamental band gap (1.12 eV \cite{vurgaftman2001band}) and the zero-point renormalization energy (0.06 eV \cite{cardona_thewalt_2005}).} & & 3.8 \footnote{Sum of the experimental room temperature optical band gap (3.53 eV \cite{tsoi2006isotopic}), the exciton binding energy (0.06 eV \cite{sun2002enhancement,fiedler2020correlative}), and a vibrational renormalization energy (0.19 eV \cite{park2022applicability}).} & & 15.3 \footnote{Sum of the experimental room temperature fundamental band gap (14.20 eV \cite{piacentini_1975}) and the zero-point renormalization energy (1.15 eV \cite{chen_pasquarello_2018, nery_gonze_2018}).} & \\
  \end{tabular}\end{ruledtabular}
\end{table}

Three additional comments are in order. First, the principles underlying our theory extend beyond the WOT approach, and in principle are relevant to aspects underlying other methods that employ removal of localized charges, several of which have been recently suggested \cite{chan_ceder_2010,nguyen_marzari_2018,degennaro2022bloch,linscott2023koopmans, ma_wang_2016,liLocalizedOrbitalScaling2018, mahler2022localized, miceli_pasquarello_2018, chen_pasquarello_2018, yang2022one}. Specifically, they allow for a meaningful connection to orbital-by-orbital-corrections based approaches \cite{borghiKoopmanscompliantFunctionalsTheir2014, imamura2011linearity}: A link is found between PWL with respect to the occupancy of a unitarily transformed orbital and PWL in charge removal from core (G)KS orbitals. See the SM \cite{SM} for further details and discussion.

Second, the present work used a specific ensemble DFT formalism to generalize the IP theorem and OT strategy to periodic systems.
Recent work has explored IP-like relations for excited state prediction~\cite{Gould2022-HL,Amoyal2024, yang_fan_2024_fractional, yang_fan_2024_orbital}.
These results suggest that OT strategies might be extended to excited state modeling, by combining recent developments in excited state ensemble DFT~\cite{Gould2017-Limits,Gould2019-DD,Gould2020-FDT,Gould2021-EGKS} and (W)OT theory. Work along these lines should be pursued.

Third, a recent study \cite{ohad_2024_chains} revealed that WOT and OT agree remarkably well also for small molecules, even though the ground-state energy of the $(N-1)$-electron system differs significantly from that of the constrained $(N-1)$-electron system. This suggests that WOT can be useful as a generalized optimal tuning strategy, beyond the solid-state limit.

To conclude, we have formulated a generalized IP condition, which relates the total energy difference upon removing an electron from a localized orbital, obtained from a unitary transformation, to its expectation value with respect to the (G)KS Hamiltonian.
This result is exact under modest assumptions, and is founded on a generalized local removal PWL condition and a generalized localized-orbital-occupation based Janak's theorem. The theory has been complemented by numerical evidence for the PWL of the total energy as a function of the fractional charge removed from the maximal-energy Wannier orbital. These insights lay a rigorous foundation for employing IP-theorem based criteria involving localized orbitals, hitherto used without proof.

This work was supported via U.S.-Israel National Science Foundation Binational Science Foundation (NSF-BSF) Grant No.~DMR-2015991 and by the Israel Science Foundation. A.R. gratefully acknowledges support from NSF-BSF Grant No.~2150562. M.C.-G. is grateful to the Azrieli Foundation for the award of an Azrieli International Postdoctoral Fellowship. T.G. and L.K. were supported by an Australian Research Council (ARC) Discovery Project (DP200100033). T.G. was supported by an ARC Future Fellowship (FT210100663). L.K. was additionally supported by the Aryeh and Mintzi Katzman Professorial Chair and the Helen and Martin Kimmel Award for Innovative Investigation.

\section*{End matter}

\subsection*{Appendix A: Physical arguments for localized charge removal}

\newcommand{\IP}{\text{IP}}

Our main proof of the localized IP theorem invoked a physically-motivated assumption about `hole' localization. Key elements are discussed in the main text. Here, we add some further details to show that the physical assumptions are, in fact, quite modest.

Consider a cube of side-length $L$ of a crystalline material.
For a sufficiently large $L$, the energy to remove a single electron will be essentially indistinguishable from the energy to remove a single electron from an infinite crystal.
Thus, we may write $\IP_L=\IP$ where the equality indicates agreement in any practicable sense.
Then, split the cube into eight smaller cubes of side-length $L/2$.
It follows that $\IP_{L/2}=\IP_L+\eta_{L/2}=\IP+\eta_{L/2}$ where $\eta_{L/2}\ll \IP$ is the change in energy associated with the smaller size.
We may repeat this procedure to obtain $\IP_{L/2^p}=\IP+\sum_{q=1}^p \eta_{L/2^q}\equiv \IP + \zeta_p$; and terminate at some $p^*$ when $\zeta_{p^*+1}\ll \IP$ is no longer satisfied.

We can now reassemble the original cube from the $8^p$ smaller cubes.
The wavefunctions associated with removal from any smaller cube may be combined with the charge-neutral wavefunctions from all other cubes to form a wavefunction for the entire cube.
The energy of this wavefunction is $\IP+\zeta_{p^*}$ and its Fukui function (i.e. the density of its  hole) is localized to the smaller cube, by construction.
Note that here we have assumed that any artefacts from combining the wavefunctions of different cubes are small -- a result that follows from the ``nearsightedness''~\cite{Kohn1996-NS} assumption of typical electronic systems, applied to small cubes far away from the edge of the large cube.
We may thus obtain degenerate wavefunctions, each confined to a cube of side-length $L/2^{p^*}$ and each with an energy within $\zeta_{p^*}\ll\IP$ of the true IP.
We may also, by symmetry, translate these wavefunctions along any integer combination of crystal lattice vectors to obtain localized states centered on any of the $M$ crystal lattice points (up to the edge effects discussed above). 
Note that these wavefunctions are not necessarily eigenstates of any crystal Hamiltonian.

Furthermore, we can minimize edge effects by applying strict periodic boundary conditions to the cubes of various sizes, so that orbital- and wavefunction-like properties obey $\phi(x+P)=\phi(x)$ where $P=L/2^p$ is the appropriate periodicity.
Then, the repeated division (or assembly) is equivalent to restricting (or loosening) the allowed periodicity of Bloch orbitals, which should obey $\phi_B(x+P)=e^{i\theta}\phi_B(x)$ in general.
We note that similar reasoning may be used to justify the restriction to a finite $\bm{k}$-grid (rather than an integral) of Bloch orbitals in periodic DFT calculations.
The mutual denominator of the $\bm{k}$-grid (on each dimension) represents the size (in crystal lattice units) of the sub-crystal captured by the Bloch orbitals and thus, in practice, imposes a strict periodicity on the solution.

The key locality assumptions of the main text may thus be reframed as follows:
(i) the smallest cube for which $\zeta_{p^*}\ll\IP$ is composed only of a small number of crystal lattice sites, so that the impact of edges is sufficiently small to be ignored (equivalent in practice to assuming that nearsightedness applies at a tens of $\text{\AA}$ scale);
(ii) the energy required to remove an electron from an excited state of any $L/2^{p^*}$-cube is significantly greater than $\zeta_{p^*}$ so that the $M$ local states are well-separated in energy from orthogonal states -- we expand on this assumption below.
It follows from these two assumptions that the wavefunctions for the smallest cubes form a complete basis for the lowest $M$ eigenstates of the original cube, and vice versa.
Using $\pazocal{U}_{pi}$ to indicate the transformation from the small cube basis to the full cube then yields
\begin{equation}
\begin{split}
\begin{aligned}
    \ket{\Psi_i^{N-1}} \equiv& \sum_{p=1}^M \pazocal{U}_{pi} \ket{\tilde{\Psi}_p^{N-1}} \\
     \Longrightarrow
    \ket{\tilde{\Psi}_p^{N-1}} \equiv& \sum_{i=1}^M \pazocal{U}_{ip} \ket{\Psi_i^{N-1}},
\end{aligned} 
\end{split}
\end{equation}
where $\pazocal{U}_{ip}=\pazocal{U}_{pi}^{*}$ follows from the properties of unitary matrices. This relation justifies the results of the main text.

Before proceeding, we note that the (a) of the main text, namely the  localization of the many-body states, may be inapplicable in systems with small or zero band gap.
The reason is that `excitations' of the lowest energy superpositions of cubes can potentially be lower in energy than the `ground states' of some of the higher energy superpositions of cubes.
A similar issue can occur for insulators, where the problematic excitations involve several different ``orbital levels'' -- i.e. a valence band.
These `excited states' can intrude on the ensemble; and thus the localized superpositions can lose some of their good properties.
However, in such cases it may be possible to form a basis of $N_b$ eigenstates on each of the $M$ lattice sites, by using the lowest $N_b$ excitations of localized states.
Here, $N_b$ must be chosen to ensure that (b) is obeyed collectively for the lowest $N_b$ states (with appropriate adjustment to $\zeta_{p^*}$).
The theory results of the main text may then be obtained from the $M\times N_b$-state system with little modification.
We note that materials like InSb and InAs, with a very small band gap, may fall under this extended definition, and that WOT works very effectively for these materials, as reported in Ref.~\cite{wing_2021}.

\subsection*{Appendix B: Finding states with equal energies}

\newcommand{\Et}{\tilde{E}}
Our central argument relies on a set of orthogonal states that (a) can be localized and (b) have the same energy as the ensemble.
(a) relies on modest physical assumptions discussed in the previous appendix.
(b) is equivalent to finding a unitary transformation $Q$ of a diagonal matrix $H$ (i.e. the matrix of the lowest $M$ eigenstates of $\hat{H}$) that makes all diagonal elements of $H_Q = Q^T H Q$ the same (since the ensemble average is $M^{-1}\Tr[H]=M^{-1}\Tr[H_Q]=M^{-1}\sum_{i=1}^M [H_Q]_{ii}$).
Conveniently, we can \emph{always} find such a transformation.

Algorithmically this is done as follows:
\begin{enumerate}
    \item Set $s=0$ and $Q_{s=0}=I$ so that $H_{Q_0}=H$ is a diagonal matrix and compute $\Et=\tfrac{1}{M}\Tr[H]$;
    \item Pick the index of the highest $h$ and lowest $l$ value entries on the diagonal of $H_{Q_s}$ (at random if degenerate):
    \begin{itemize}
    \item Terminate and return $H_{Q_s}$ if $|[H_{Q_s}]_{hh}-\Et|<\eta$ and $|[H_{Q_s}]_{ll}-\Et|<\eta$ for some infinitesimal $\eta$;
    \end{itemize}
    \item Define a rotation matrix $R$ on $h$ and $l$ with angle:
    \begin{itemize}
        \item $\theta=\pm\sin^{-1}\sqrt{\tfrac{[H_{Q_s}]_{hh}-\Et}{[H_{Q_s}]_{hh}-[H_{Q_s}]_{ll}}}$ if $|[H_{Q_s}]_{hh}-\Et|>|[H_{Q_s}]_{ll}-\Et|$, so that $[R^TH_{Q_s}R]_{hh}=\Et$;
        \item $\theta=\pm\sin^{-1}\sqrt{\tfrac{\Et-[H_{Q_s}]_{ll}}{[H_{Q_s}]_{hh}-[H_{Q_s}]_{ll}}}$ otherwise, so that $[R^TH_{Q_s}R]_{ll}=\Et$;
        \item (the sign may be chosen at random)
    \end{itemize}
    \item Set $H_{Q_{s+1}}=R^TH_{Q_s}R$, $Q_{s+1}=Q_sR$ and $s\to s+1$ and repeat from Step 2.
\end{enumerate}
Note that selecting the extremes for $h_s$ and $l_s$ -- where subscript $s$ indicates that these indices change at each step -- ensures that the unitary transformation is applied only once to each pair of terms.
This is because one or the other term becomes equal to to $\Et$, and thus cannot contribute to future iterations.
Consequently, the off-diagonal elements $[H_{Q_s}]_{h_sl_s}=[H_{Q_s}]_{l_sh_s}$ (initially zero) become non-zero only \emph{after} $h_s$ or $l_s$ is eliminated and so do not interfere with Step~3.
It follows from the invariance of the trace that $S\leq M$ repetitions must ensure that \emph{all} diagonal elements are equal to $\Et$, since each step adds at least one additional $\Et$ to the diagonal.

This result reveals that (b) is guaranteed provided $M$ is finite and the spectrum of $H$ is bounded above and below—conditions typically satisfied by physical systems.
It is thus only its combination with (a) that requires physical motivation.

\bibliographystyle{apsrev4-2.bst}
\bibliography{references.bib}

\end{document}


\pagestyle{plain}

\title{Supplemental Material: Foundations of the ionization potential condition for localized electron removal in density functional theory}

\author{Guy Ohad}
    \affiliation{Department of Molecular Chemistry and Materials Science, Weizmann Institute of Science, Rehovoth 76100001, Israel}
\author{Mar\'ia Camarasa-G\'omez}
    \affiliation{Department of Molecular Chemistry and Materials Science, Weizmann Institute of Science, Rehovoth 76100001, Israel}
    \affiliation{Centro de F\'isica de Materiales (CFM-MPC), CSIC-UPV/EHU, Paseo Manuel de Lardizabal 5, 20018 Donostia-San Sebastián, Spain}
\author{Jeffrey B. Neaton}
    \affiliation{Department of Physics, University of California, Berkeley, Berkeley, California 94720, USA}
    \affiliation{Materials Sciences Division, Lawrence Berkeley National Laboratory, Berkeley, California 94720, USA}
    \affiliation{Kavli Energy NanoSciences Institute at Berkeley, University of California, Berkeley, Berkeley, California 94720, USA}
\author{Ashwin Ramasubramaniam}
    \affiliation{Department of Mechanical and Industrial Engineering, University of Massachusetts Amherst, Amherst, MA 01003, USA}
    \affiliation{Materials Science and Engineering Graduate Program, University of Massachusetts Amherst, Amherst, MA 01003, USA}
\author{Tim Gould}
    \affiliation{Queensland Micro- and Nanotechnology Centre, Griffith University, Nathan, QLD 4111, Australia}
\author{Leeor Kronik}
   \affiliation{Department of Molecular Chemistry and Materials Science, Weizmann Institute of Science, Rehovoth 76100001, Israel}

\maketitle
\thispagestyle{plain}

\section{Connection of the theory to orbital-by-orbital correction schemes}

We discuss an additional perspective on PWL of the total energy with respect to Wannier occupancies. In light of the chain rule employed in Eq.~(21) of the main text, we can rewrite result (22) of the main text as
\begin{equation}
    \frac{\partial E}{\partial\tilde{f}_{ii'}} = \sum_{j}U_{ji}U_{ji'}^{*}\epsilon_{j}(f_j),
\end{equation}
where $\epsilon_{j}(f_j)$ emphasizes that the eigenvalues solely depend on their corresponding occupancy, as the partial derivative $\frac{\partial E}{\partial f_j}$ assumes that $f_j$ is the only occupancy allowed to vary. Hence, the total energy is piecewise linear with respect to $\tilde{f}_{ii'}$ if $\epsilon_{j}$ are constant with respect $f_j$. Whether this condition is true or not for the exact functional is unknown, to the best of our knowledge, as non-Aufbau variations are generally not well defined within DFT. Nonetheless, this condition has been employed successfully and shown to improve electronic structure predictions \cite{borghiKoopmanscompliantFunctionalsTheir2014, imamura2011linearity, linscott2023koopmans}.

\begin{figure}[htbp!]
\begin{centering}
\includegraphics[width=0.75\linewidth]{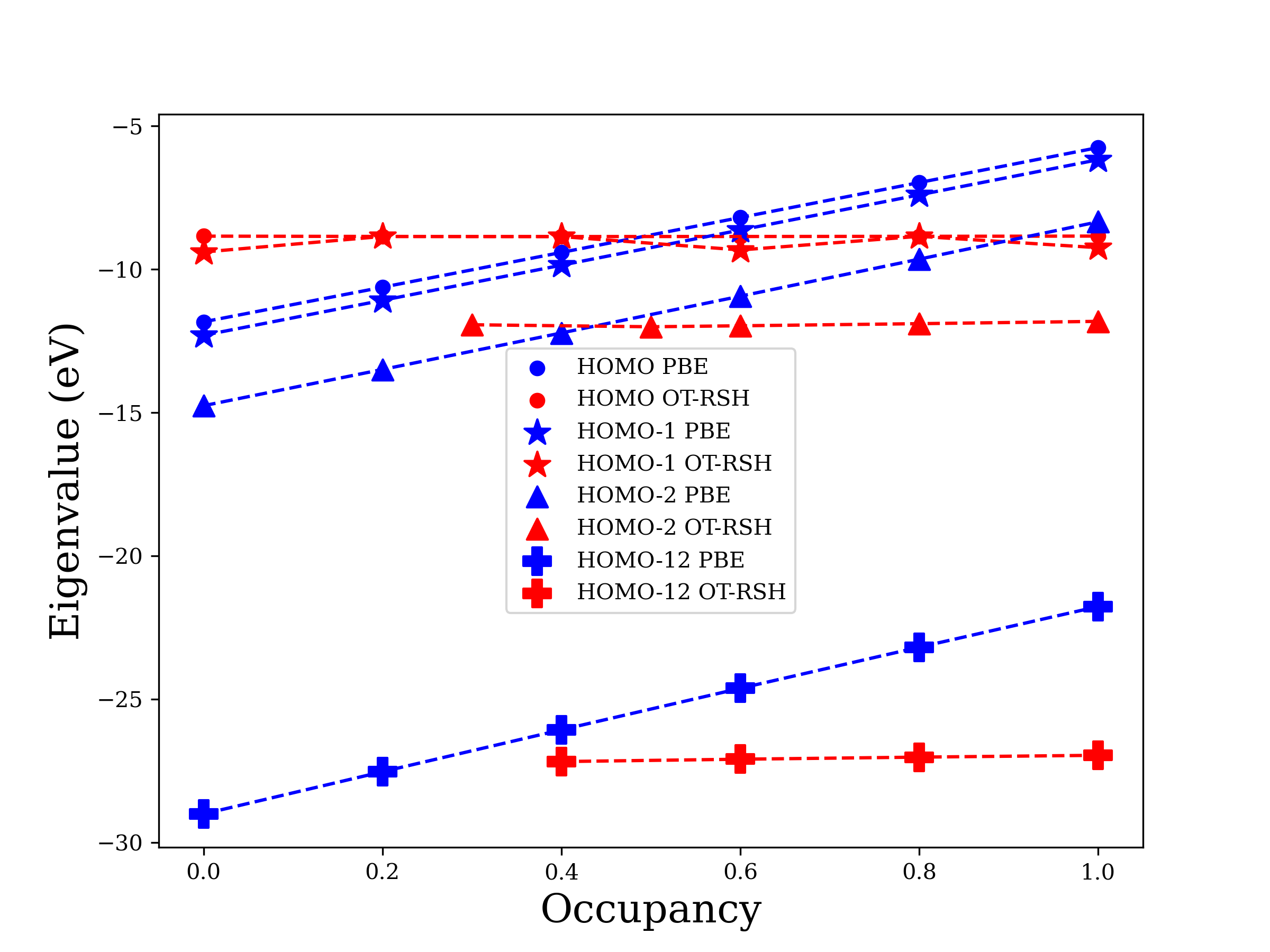}
    \caption{Eigenvalues corresponding to the HOMO minus $n$ orbital for $n=0,1,2,12$ as a function of their occupancy in thiophene, using PBE (blue) and OT-RSH (red). Each curve represents a set of calculations where the occupancy of the HOMO minus $n$ orbital is allowed to change while the other orbital occupancies are kept fixed. The slopes of the curves, in eV, are: PBE: 6.08, 6.13, 6.42, and 7.24; OT-RSH: 0.00, 0.04, 0.21, and 0.36, for $n=0,1,2,12$ respectively. All calculations are performed with the \textsc{NWChem} software package \cite{apra2020nwchem}, using the cc-pvtz basis set.}
    \label{fig:eigenvalue_vs_occ}
\end{centering}
\end{figure}

To explore the validity of this condition, we examine occupancy variations for core electrons of the thiophene molecule using PBE and OT-RSH. Fig.~\ref{fig:eigenvalue_vs_occ} shows variations in the eigenvalues of the highest occupied molecular orbital (HOMO) and selected core orbitals with respect to their occupancy. By design, the HOMO eigenvalue in OT-RSH is constant with its occupancy, while the HOMO eigenvalue in PBE changes substantially. This demonstrates the known (deviation from) PWL of these DFAs. Lower lying eigenvalues in both OT-RSH and PBE show stronger variations with respect to their occupancy, indicated by their slope. But in OT-RSH the slopes are still smaller by an order of magnitude compared to the corresponding slopes of the PBE eigenvalues. The fact that even core eigenvalues in OT-RSH are roughly constant with their occupancy, despite not being strictly constrained to be, demonstrates an interesting property of core eigenvalues when the HOMO is constrained to remain constant with its occupancy. Because the HOMO eigenvalue in the exact functional is strictly constant, this indicates on a possible property of the core eigenvalues being approximately constant as well in the exact functional. The fact that the degree of constancy decreases when going down to deeper core eigenvalues is a possible indication for the importance of constructing the Wannier orbital from the top of the valence manifold and choosing the one with highest $\tilde{\epsilon}_{ii}$, as done in the WOT procedure. Additionally, it may suggest that the constancy condition deteriorates for deep levels. These trends are consistent with previous findings on the relation between lower-lying occupied KS eigenvalues and exact core-level ionization energies: although no formal connection exists, the highest occupied KS eigenvalues often approximate the exact quasiparticle removal energies well, with discrepancies increasing for deeper-lying states \cite{kronik_kuemmel_2014, jones_gunnarsson_1989}.

To conclude, this perspective establishes a connection between WOT and the generalized linearity condition for all eigenvalues \cite{borghiKoopmanscompliantFunctionalsTheir2014, imamura2011linearity, linscott2023koopmans} and highlights the underlying link between the success of these methods \cite{borghiKoopmanscompliantFunctionalsTheir2014, imamura2011linearity, linscott2023koopmans} in accurately predicting electronic and optical excitations.

\section{Numerical procedure for fractional Wannier occupancy calculations}

In order to inspect PWL numerically, we have to define a procedure for removing a fraction of an electron from a Wannier orbital in a gapped system. In a system with $N$ electrons, the density and kinetic energy can be written as
\begin{equation}
    \begin{aligned}
        n^N(\bm{r}) & = \sum_{i=1}^N \abs{\psi_i^N(\bm{r})}^2 = \sum_{i=1}^N \abs{\tilde{\psi}_i^N(\bm{r})}^2\\
        T_s^N & =  \sum_{i=1}^N \bra{\psi_i^N}-\tfrac{1}{2}\nabla^2\ket{\psi_i^N} =  \sum_{i=1}^N \bra{\tilde{\psi}_i^N}-\tfrac{1}{2}\nabla^2\ket{\tilde{\psi}_i^N}.
    \end{aligned}
\end{equation}
Importantly, in this case a unitary transform is applied to the occupied manifold while unoccupied (and possibly core) orbitals are left unchanged (the overall transformation is still unitary). 


To ensure that a fractional charge is removed from a Wannier orbital, we demand that in a system with $N-q$ electrons \cite{ma_wang_2016}
\begin{equation}
    \begin{aligned}
    \label{eq:frac_wann_density_kinetic}
        n^{N-q} (\bm{r})  = & \sum_{i=1}^{N-1} \abs{\tilde{\psi}_i^{N-q}(\bm{r})}^2 + (1-q) \abs{\tilde{\psi}_H^{N-q}(\bm{r})}^2\\
        T_s^{N-q} = & \sum_{i=1}^{N-1} \bra{\tilde{\psi}_i^{N-q}}-\tfrac{1}{2}\nabla^2\ket{\tilde{\psi}_i^{N-q}} 
         + (1-q) \bra{\tilde{\psi}_H^{N-q}}-\tfrac{1}{2}\nabla^2\ket{\tilde{\psi}_H^{N-q}},
    \end{aligned}
\end{equation}
where $\tilde{\psi}_H$ is the Wannier orbital associated with the maximal $\tilde{\epsilon}_{ii}$. In order to simplify the procedure, we assume that $\tilde{\psi}_H$ does not relax much upon fractional electron removal, such that for every $q$, $\tilde{\psi}_H^{N-q} \approx \tilde{\psi}_H^N$. The total energy of the fractional Wannier system is computed by fixing the Wannier occupancies according to $\tilde{f}_{ii} = 1-q\delta_{iH}$. The orbitals $\qty{\tilde{\psi}_i}_{i=1}^{N-1}$ are then allowed to relax while being orthogonal to $\tilde{\psi}_H^N$ via an energy penalty method, namely
\begin{equation}
\label{eq:constr_frac_wann_total_energy}
    \begin{aligned}
    \tilde{E}^{N-q}= & \min_{\{\tilde{\psi}_i\}}\ \left[ E^{N-q}[\{\tilde{\psi}_i\}]
     + \frac{1}{2} \sum_j  \lambda_j \left( \abs{\bra{\tilde{\psi}_H^N}\ket{\tilde{\psi}_j}}^2 - \delta_{jH} \right)^2 \right],
    \end{aligned}
\end{equation}
where $\lambda_j$ are Lagrange multipliers, $E^{N-q}[\{\tilde{\psi}_i\}]$ is calculated with the density and kinetic energy in Eq.~\eqref{eq:frac_wann_density_kinetic}, and $\tilde{E}^{N-q}$ already includes image-charge corrections \cite{makov_payne_1995,leslie_gillan_1985, komsa_pasquarello_2012}. Taking the functional derivative of Eq.~\eqref{eq:constr_frac_wann_total_energy} then yields the equations \begin{equation}   \hat{H}_{\textrm{KS}}^{N-q} \ket{\tilde{\psi}_i}+  \sum_j \lambda_j \left( \abs{\bra{\tilde{\psi}_H^N}\ket{\tilde{\psi}_j}}^2 - \delta_{jH} \right) \ket{\tilde{\psi}_H^N} \bra{\tilde{\psi}_H^N}  \ket{\tilde{\psi}_i}  = \epsilon^{'}_{i} \ket{\tilde{\psi}_i},\end{equation} which are solved self-consistently using a conjugated gradient procedure. $\qty{\lambda_j}$ should be selected to be large enough to enforce the desired orthogonality. 

\section{Computational details}

All calculations were performed with an in-house modified version of the Vienna \textit{ab initio} simulation package (\textsc{VASP}) \cite{Kresse_1996}. We use the PBE-based projector augmented wave (PAW) method for treating core electrons \cite{Kresse_1999}. The valence configuration included in the PAWs is $2s^22p^63s^23p^2$ for Si, $3d^{10}4s^2$ for Zn, $2s^22p^4$ for O, $2s^1$ for Li, and $2s^22p^5$ for F. We made use of the \textsc{wannier90} software package \cite{mostofi_marzari_2014} for generating maximally localized Wannier orbitals. Spin-orbit coupling effects were neglected throughout.

For the fractional Wannier occupation calculations, we use $2\times2\times2$ cubic supercells evaluated with a $\Gamma$-point-only sampling of the Brillouin zone. We enforce the occupation of the Wannier orbital by setting $\lambda_j=300$ eV. To correct the energy of the $(N-q)$-electron system we use the Makov-Payne image charge correction \cite{makov_payne_1995,leslie_gillan_1985, komsa_pasquarello_2012}, given by
\begin{equation}
    \Delta E = \frac{\alpha_m q^2}{2\varepsilon_\infty L},
\end{equation}
where $\alpha_m=2.837$ is the the Madelung constant for a cubic cell, $L$ is the length of the supercell, and $\varepsilon_\infty$ is the orientationally averaged ion-clamped dielectric tensor of the material. We select the $\varepsilon_\infty$ value obtained in the WOT-SRSH procedure (see below).

The SRSH functional employs three parameters: The fraction of short-range exact (Fock) exchange, $\alpha$, the fraction of long-range exact exchange, $\varepsilon_\infty^{-1}$, and the range-separation parameter, $\gamma$ \cite{refaely-abramson_kronik_2013, wing_2021}. For the materials studied here, we used the WOT-SRSH parameters determined in previous studies \cite{wing_2021, ohad_gant_2023_metal_oxides}, with slight modifications owing to the use of other PAWs and/or plane-wave code. The final parameters are given in Table \ref{tab:srsh_params}. We additionally selected non-optimal SRSH parameters such that the band gap is overestimated (SRSH$\uparrow$) and underestimated (SRSH$\downarrow$) with respect to the WOT-SRSH band gap. These parameters are also given in Table \ref{tab:srsh_params}.

\begin{table}[htbp!]
  \centering
  \caption{Functional parameters for WOT-SRSH, SRSH$\uparrow$, and SRSH$\downarrow$.}
  \label{tab:srsh_params}
  \begin{tabular}{|c|ccc|ccc|ccc|}
    \hline
    & \multicolumn{3}{|c|}{Si} & \multicolumn{3}{|c|}{ZnO} & \multicolumn{3}{|c|}{LiF} \\
    \hline
        & $\alpha$ & $\varepsilon_\infty^{-1}$ & $\gamma$ (\AA$^{-1}$) & $\alpha$ & $\varepsilon_\infty^{-1}$ & $\gamma$ (\AA$^{-1}$) & $\alpha$ & $\varepsilon_\infty^{-1}$ & $\gamma$ (\AA$^{-1}$) \\
    \hline
    WOT-SRSH & 0.25 & 0.08 & 0.38 & 0.30 & 0.28 & 1.30 & 0.25 & 0.52 & 1.95 \\
    \hline
    SRSH$\uparrow$ & 0.90 & 0.08 & 0.15 & 0.30 & 0.60 & 0.60 & 0.25 & 0.80 & 1.20 \\
    \hline
    SRSH$\downarrow$ & 0.15 & 0.08 & 1.00 & 0.10 & 0.28 & 0.20 & 0.25 & 0.52 & 0.20 \\
    \hline
  \end{tabular}
\end{table}

\bibliographystyle{apsrev4-2.bst}
\bibliography{references.bib}